\documentclass[12pt]{article}

\begin{document}


\newcommand{\bb}{\begin{equation}}
\newcommand{\ee}{\end{equation}}
\newcommand{\bbb}{\begin{eqnarray}}
\newcommand{\eee}{\end{eqnarray}}
\newcommand{\vc}[1]{\mbox{$\vec{{\bf #1}}$}}
\newcommand{\mc}[1]{\mathcal{#1}}
\newcommand{\del}{\partial}
\newcommand{\lk}{\left}
\newcommand{\ave}[1]{\mbox{$\langle{#1}\rangle$}}
\newcommand{\re}{\right}
\newcommand{\pd}[1]{\frac{\del}{\del #1}}
\newcommand{\pdd}[2]{\frac{\del^2}{\del #1 \del #2}}
\newcommand{\Dd}[1]{\frac{d}{d #1}}
\newcommand{\sech}{\mbox{sech}}
\newcommand{\pref}[1]{(\ref{#1})}

\newcommand
{\sect}[1]{\vspace{20pt}{\LARGE}\noindent
{\bf #1:}}
\newcommand
{\subsect}[1]{\vspace{20pt}\hspace*{10pt}{\Large{$\bullet$}}\mbox{ }
{\bf #1}}
\newcommand
{\subsubsect}[1]{\hspace*{20pt}{\large{$\bullet$}}\mbox{ }
{\bf #1}}

\def\ie{{\it i.e.}}
\def\eg{{\it e.g.}}
\def\cf{{\it c.f.}}
\def\etal{{\it et.al.}}
\def\etc{{\it etc.}}

\def\AA{{\cal A}}
\def\BB{{\cal B}}
\def\CC{{\cal C}}
\def\DD{{\cal D}}
\def\EE{{\cal E}}
\def\FF{{\cal F}}
\def\GG{{\cal G}}
\def\HH{{\cal H}}
\def\II{{\cal I}}
\def\JJ{{\cal J}}
\def\KK{{\cal K}}
\def\LL{{\cal L}}
\def\MM{{\cal M}}
\def\NN{{\cal N}}
\def\OO{{\cal O}}
\def\PP{{\cal P}}
\def\QQ{{\cal Q}}
\def\RR{{\cal R}}
\def\SS{{\cal S}}
\def\TT{{\cal T}}
\def\UU{{\cal U}}
\def\VV{{\cal V}}
\def\WW{{\cal W}}
\def\XX{{\cal X}}
\def\YY{{\cal Y}}
\def\ZZ{{\cal Z}}

\def\sinh{{\rm sinh}}
\def\cosh{{\rm cosh}}
\def\tanh{{\rm tanh}}
\def\sgn{{\rm sgn}}
\def\det{{\rm det}}
\def\trace{{\rm Tr}}
\def\exp{{\rm exp}}
\def\sh{{\rm sh}}
\def\ch{{\rm ch}}

\def\ell{{\it l}}
\def\str{{\it str}}
\def\lp{\ell_{{\rm pl}}}
\def\blp{\overline{\ell}_{{\rm pl}}}
\def\ls{\ell_{{\str}}}
\def\bls{{\bar\ell}_{{\str}}}
\def\bM{{\overline{\rm M}}}
\def\gs{g_\str}
\def\gym{{g_{Y}}}
\def\geff{g_{\rm eff}}
\def\eff{{\rm eff}}
\def\r11{R_{11}}
\def\kel{{2\kappa_{11}^2}}
\def\kten{{2\kappa_{10}^2}}
\def\lpten{{\lp^{(10)}}}
\def\alp{{\alpha'}}
\def\aleff{{\alp_{eff}}}
\def\sqaleff{{\alp_{eff}^2}}
\def\tgs{{\tilde{g}_s}}
\def\talp{{{\tilde{\alpha}}'}}
\def\tlp{{\tilde{\ell}_{{\rm pl}}}}
\def\tr11{{\tilde{R}_{11}}}
\def\wtilde{\widetilde}
\def\what{\widehat}
\def\hlp{{\hat{\ell}_{{\rm pl}}}}
\def\hr11{{\hat{R}_{11}}}
\def\hf{{\textstyle\frac12}}
\def\coeff#1#2{{\textstyle{#1\over#2}}}
\def\CY{Calabi-Yau}
\def\lessapprox{\;{\buildrel{<}\over{\scriptstyle\sim}}\;}
\def\greaterapprox{\;{\buildrel{>}\over{\scriptstyle\sim}}\;}
\def\inbar{\,\vrule height1.5ex width.4pt depth0pt}
\def\IC{\relax\hbox{$\inbar\kern-.3em{\rm C}$}}
\def\IR{\relax{\rm I\kern-.18em R}}
\def\IP{\relax{\rm I\kern-.18em P}}
\def\Z{{\bf Z}}
\def\R{{\bf R}}
\def\One{{1\hskip -3pt {\rm l}}}
\def\sst{\scriptscriptstyle}
\def\osc{{\rm\sst osc}}
\def\lam{\lambda}
\def\lc{{\sst LC}}
\def\pr{{\sst \rm pr}}
\def\cl{{\sst \rm cl}}
\def\D{{\sst D}}
\def\bh{{\sst BH}}
\def\vev#1{\langle#1\rangle}

\input{epsf}

\begin{titlepage}
\rightline{CLNS 99/1639}

\rightline{hep-th/9910099}

\vskip 1cm
\begin{center}
\Large{{\bf 
Holography, a covariant c-function\\
and the geometry of the renormalization group
}}
\end{center}

\vskip 1cm
\begin{center}
Vatche Sahakian\footnote{\texttt{vvs@mail.lns.cornell.edu}}
\end{center}
\vskip 12pt
\centerline{\sl Laboratory of Nuclear Studies}
\centerline{\sl Cornell University}
\centerline{\sl Ithaca, NY 14853, USA}

\vskip 2cm

\begin{abstract}
We propose a covariant geometrical expression for the c-function 
for theories which admit
dual gravitational descriptions. We state a c-theorem 
with respect to this quantity
and prove it. We apply the expression to a class of geometries,
from domain walls in gauged supergravities, to extremal and near
extremal $Dp$ branes, and the AdS Schwarzschild black hole. 
In all cases, we find agreement with expectations.

\end{abstract}

\end{titlepage}
\newpage
\setcounter{page}{1}

\section{Introduction}

The holographic encoding of
information in gravitational theories appears to be a manifestation of
a fundamental physical principle. The importance of this
projection of information was realized in the context of classical
general relativity through entropy bounds 
and black hole thermodynamics~\cite{BEKEN1,BEKEN2,HAWKING1}.
More recently, 
we have learned from string theory that this phenomenon appears to have
intriguing connections with scaling and
renormalization group flow 
in non-gravitational theories~\cite{GPPZ,DISTZAM,KEHSFET,PORRSTAR,GPPZ2,
BALARG,FGPW,SKENTOWN}.
A great deal remains to be unraveled about this connection, but there
are already indications that this line of thought may hold the resolution of
some of the paradoxical issues arising from black hole physics.

In this work, we investigate the connection between renormalization
group flow and holography by proposing a covariant,
geometric measure for the effective central charge
for the so called 
``boundary theory'' of Maldacena's duality~\cite{MALDA1,WITHOLO,GUBSER}.
Central charge, or the ``c-function'', is a measure of
the degrees of freedom of a theory, the number of independent
species of excitations. For theories in two dimensions,
Zamolodchikov~\cite{ZAMO} was able to prove a set of 
elegant statements describing
the central role played by this quantity in the renormalization group flow.
The effective central charge was shown to be
a function of the couplings of the theory that is
monotonically decreasing as one flows to lower
energies; fixed points described by conformal field theories
correspond to the extrema of this function, and
the gradients over coupling space are related to the beta functions
of the theory. Attempts to generalize some of these statements
to higher dimensions have been met with very limited success.
In the context of Maldacena's duality, we acquire geometrical tools
to study this question in regimes where a theory is strongly coupled.

The basic conceptual ingredient in our proposal is a remarkably simple,
yet powerful prescription proposed by Bousso~\cite{BOUSSOCONJ}. The 
observation is that holographic statements
should have a covariant nature. Consequently, Bousso
proposes to use congruences of null
geodesics as probes for the sampling of holographically encoded information.
We believe that this principle is a general one. The proposal of 
Maldacena in regimes where one focuses on bulk dynamics in
the non-stringy gravity sector must have
a similar covariant nature. 

In the next section, we motivate and construct a covariant expression
for the central charge for the boundary theory.
We will guide ourselves by a set of intuitively driven principles inspired
by the Bousso entropy bound. We will then prove
a c-theorem; Bousso's criterion
for the convergence of the congruence, along with 
the null convergence criterion, are identified as the necessary and
sufficient conditions 
\footnote{This is for cases involving shearless flow; a more
general statement can also be proved for the cases with shear.}.
We will then proceed to apply the prescription 
to certain classes of geometries:
domain wall solutions in gauged supergravities, near horizon regions of
extremal and near-extremal $Dp$ branes for $p<5$, and the 
Anti-de Sitter (AdS) Schwarzschild
black hole in four dimensions. 
In the first class, we find exact agreement with~\cite{FGPW}.
For the second class, the scaling of the
c-function is found to match onto the expected
asymptotic behaviors given by the perturbative
Supersymmetric Yang-Mills (SYM),
the Matrix String, the M theory membrane and the M theory
five brane theories. We also find that
our expression, applied to flow
along ``radial congruences'', is insensitive to the presence of 
a thermodynamic horizon; as expected, the latter corresponds to
a thermodynamic state in the same dual theory.
We end with a discussion assessing the evidence presented.

\section{A covariant c-function}

Consider a $D$ dimensional spacetime with metric $g_{ab}$
foliated by a choice of
constant time surfaces. Let this vacuum solve Einstein's equations
with a negative cosmological constant.
Focus on a spacelike $D-2$ dimensional surface $\MM$
at some fixed time. There are
generally four light-sheets projected out of this surface consisting
of the spacetime points visited by a congruence of null geodesics orthogonal
to $\MM$~\cite{WALD}. 
As prescribed by Bousso~\cite{BOUSSOCONJ}, we pick a light-sheet
along which the congruence converges. If $n^a$ denotes
the tangents to these geodesics, we can construct a null vector field $m^a$
on the light-sheet such that it is orthogonal to $\MM$ and satisfies
$m^a n_a = -1$~\cite{FLAN}. Then $\MM$ admits the metric
\bb
h_{ab}=g_{ab}+n_{(a} m_{b)}\ .
\ee
The {\em second fundamental form} is defined by~\cite{WALD,HAWKELLIS}
\bb
B_{ab}\equiv \nabla_b n_a= \nabla_a n_b = B_{ba}\ .
\ee
The symmetry property follows from Frobenius's theorem and
the fact that the vector field $n^a$ is surface orthogonal.
We define the $D-2$ dimensional matrix
\bb
{\hat{B}}^a_b=g^{ac} B_{cb}=h^{ac} B_{cb}\ ,
\ee
and its trace
\bb
\theta=\trace\ \hat{B}\ .
\ee
The condition for the convergence of the 
geodesics is stated as~\cite{BOUSSOCONJ}
\bb
\theta \le 0\ .
\ee
The geodesics are to be extended as long as this condition is satisfied;
for spacetimes curved by matter satisfying the null convergence
condition\footnote{The null convergence condition is discussed
in detail in~\cite{WALD}; it is the statement that the energy
momentum tensor satisfies the condition $T_{ab}k^a k^b\ge 0$
for all null $k^a$. This follows from the strong or weak
energy conditions that are believed to be satisfied by all known
forms of matter.}, 
these geodesics will
typically end at caustics ($\theta\rightarrow -\infty$).
We would like to think of a sense in which information on such a light-sheet
is holographically encoded on $\MM$. This is the nature of Bousso's
entropy bound, and it is also consistent with Maldacena's
conjecture\footnote{The past/future history of the light-sheet is causally 
related to it; so that the spacelike notion of holography we may naively
accord to Maldacena's duality is a subset of this covariant 
statement.}.

For simplicity, let us require that the 
spacetimes we consider admit a timelike Killing vector
field along which our choice of time flows. 
One may propose to time flow the whole light-sheet forward
and backward, generating a D dimensional region of spacetime which becomes
the ``bulk''; and the ``boundary'' is the time flow of $\MM$. 
The c-function we will be considering
is to be accorded to the theory that is in some sense living on this boundary 
and is dual to the bulk. We want to associate 
renormalization group flow to lower energy scales with motion along the
converging congruence of null geodesics. 
A more precise and careful version of this statement 
will be postponed to future work. 
For now, we will use the null geodesics described above
as tools to probe covariantly the dual theory at lower energy scales; 
the success of our proposal
in the examples we will consider can be viewed as evidence to this approach.
\begin{figure}
\epsfxsize=11cm \centerline{\leavevmode \epsfbox{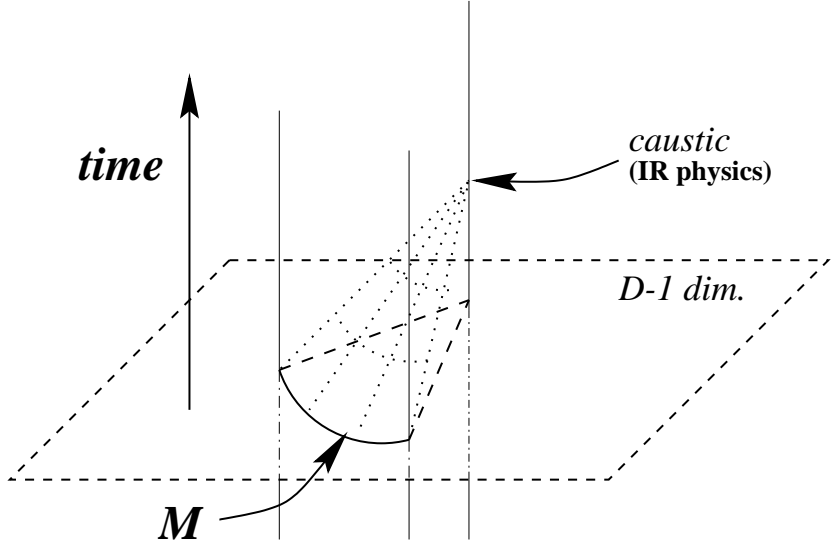}}
\caption{\sl An illustration of the construction of holographic duals;
the $D-2$ dimensional surface $\MM$ is shown, along with a caustic
ending the congruence of null geodesics.
}
\label{f1fig}
\end{figure}

Without claim to rigor, we next motivate the geometrical formula for the
c-function. The first principle we accord to is that the {\em local
geometry} transverse to the flow
encodes the information about the decrease of the effective central charge
due to the coarse graining of the boundary theory. This is 
partly motivated
by the work of~\cite{BALASTRESS}, where the energy momentum tensor of the
boundary theory for AdS spacetimes was written in terms of local
quantities, essentially the extrinsic curvature of a foliation and its trace.
From the same line of thought, we expect the c-function to be proportional
to $G_D^{-1}$, the inverse of the $D$ dimensional gravitational
coupling. The next tool is dimensional analysis; we need a local 
{\em covariant} object with dimensions ${length}^{D-2}$ to construct
a dimensionless c-function; 
motivated by the Bousso construction,
we allow ourself to use only covariant data from the congruence.
Intuitively, we also require {\em invariance} under boundary diffeomorphisms;
the most natural way to assure this is through an integration over
the boundary using the proper measure constructed from $h_{ab}$. Finally,
we require that the formula gives the proper scaling for the central
charge in AdS spaces, 
\ie\ a constant proportional to $l^{D-2}/G_D$~\cite{HENNSKEN}, where
the cosmological constant scales as $-1/l^2$. For AdS spaces with metric in
the Poincar\'{e} coordinates
\bb
ds^2=\frac{l^2}{z^2} \lk( dz^2+ d{\vec{x}}^2_{D-2} - dt^2\re)
\ee
and a choice of constant $z$ foliations at fixed time for $\MM$,
we have $\theta\sim z/l^2$ and $\sqrt{h}\sim (l/z)^{D-2}$. 
Putting everything together, we
are left with combinations of two simple expressions, $\det \hat{B}$ and
$\theta$, 
differing from each other 
only if the flow of the congruence has shear (see definition below);
we pick a form that appears the most natural:
\bb\label{prop1}
c(\tau)\simeq \frac{c_0}{G_D \int \lk.\sqrt{h}\  \det \hat{B}\re |_{\tau}}\ .
\ee
Another obvious option is to replace $(\det \hat{B})$ by $\theta^{D-2}$.
The integral is evaluated at some proper time $\tau$ along the geodesics;
$\tau$ is to be related to the energy scale of renormalization group
flow through the UV-IR relation. There are two problems
with expression~\pref{prop1}; dependence on the scale parameter for the proper
time; and it is divergent for many relevant geometries with non-compact
boundaries $\MM$.
Both problems can be regulated by proposing that the local expression 
we have can yield only the fractional decrease in the central charge as 
a result of renormalization. We then propose the cure: 
\bb\label{prop}
\frac{c_{\tau_2}}{c_{\tau_1}}=
\frac{G_D \int \lk.\sqrt{h}\ \det \hat{B}\re |_{\tau_1}}
{G_D \int \lk.\sqrt{h}\ \det \hat{B}\re |_{\tau_2}}\ .
\ee
One fixes the central charge at some proper time $\tau_1$ from other data,
and the formula predicts the central charge at lower energy scales deep
in the bulk at $\tau_2$. The expression is manifestly covariant, independent of
the arbitrary scaling of the proper time, in principle convergent, and
invariant under boundary diffeomorphisms. Typically, one would expect
to work in a gravitational theory with negative cosmological constant
so as to have an asymptotic AdS vacuum configuration corresponding
to a UV fixed point. $c_{\tau_1}$ gets fixed by the conformal algebra 
in this region, and $c_{\tau_2}$ predicts the central charge of the deformed
conformal theory at lower energy scales. The product of $c_{\tau_1}$ with
the numerator of the right hand side is a numerical coefficient times
regulator factors canceling with the denominator of the right hand side.
For pure AdS spaces, equation~\pref{prop} gives one by construction.

In practical calculations, we will have symmetries that allow us
to write a slightly simplified formula. For the rest of this work, we
assume that the flow under consideration is shearless; the matrix $\hat{B}$
can be decomposed generally as~\cite{WALD}
\bb
\hat{B}= \frac{\theta}{D-2} {\bf 1} + \hat{\sigma}\ ,
\ee
where the symmetric matrix $\hat{\sigma}$ is referred to as shear. For
all cases under considerations, we have $\hat{\sigma}=0$, so that 
$\hat{B}$ is proportional to a $D-2$ dimensional identity matrix. We then write
\bb\label{propt}
c(\tau)=\frac{c_0}{G_D \int \lk.\sqrt{h}\  \det \hat{B}\re |_{\tau}}
\rightarrow \frac{c'_0}{G_D \int \lk.\sqrt{h}\  \theta^{D-2}\re |_{\tau}}\ ,
\ee
where $c_0$ and $c'_0$ are products of numerical coefficients, and 
a factor regulating the size of $\MM$ in the cases where $\MM$ is non-compact.
As mentioned above, for shearless flow, 
we are unable to discriminate between combinations of the two
expressions that we were led to in the arguments above; \ie\ 
$\det \hat{B}\sim \theta^{D-2}$. We are however intuitively driven
to propose that the general form for the central charge should
be given by~\pref{prop}. We expect that
the flow of the central charge should be
sensitive to the phenomenon of shear.

We will next prove a c-theorem for equation~\pref{propt}.
A more general version with respect
to equation~\pref{prop} can be proven 
as well (the convergence criterion gets slightly generalized and
we would require that the congruence is
principal null with respect to the Weyl tensor).
However, our understanding of the physical role of shear from
the renormalization group perspective is primitive; we defer the
more general statement to future work where we hope to explore 
explicitly examples with shear.

\section{The c-theorem}

\paragraph{Theorem:} 
Consider a congruence of null geodesics emanating from a $D-2$
dimensional surface $\MM$ as defined above; then equation~\pref{propt}
is monotonically decreasing for increasing $\tau$ if the null convergence
condition is satisfied, and if everywhere along
the flow $\theta\le 0$.

\vspace{15pt}

The proof is straightforward.
Differentiating the log of equation~\pref{propt}, 
the derivative slices through the integral (the congruence is
orthogonal to $\MM$) and we get
\bb\label{prf1}
\frac{d}{d\tau}\ln\ c = -\frac{1}{\int \sqrt{h} \theta^{D-2}}
\int \frac{d}{d\tau} \lk( \sqrt{h} \theta^{D-2}\re)\ ;
\ee
The monotonicity follows immediately from Raychaudhuri's equation
\bb
\frac{d\theta}{d\tau}=-\frac{1}{D-2} \theta^2
-R_{ab} n^a n^b\ ,
\ee
and from 
\bb
\frac{d}{d\tau} \sqrt{h} = \delta_n \sqrt{h} = \sqrt{h}\ \theta\ .
\ee
We then have
\bb\label{derprop}
\frac{d}{d\tau} \ln\ c= - \frac{(D-2)}{\int \sqrt{h} \ |\theta|^{D-2}}
\int \sqrt{h}\ |\theta |^{D-3} R_{ab} n^a n^b \le 0\ ,
\ee
where we have used $\theta\le 0$. $R_{ab} n^a n^b \ge 0$ follows
from the null convergence condition and Einstein's equations
in the presence of a cosmological constant since $n^a n_a=0$.
The null convergence condition follows from either the weak
or strong energy conditions.

\vspace{15pt}
Note that equation~\pref{derprop} vanishes for $D=2$, 
\ie\ when the boundary is described by quantum
mechanics. This is consistent with the fact that
the renormalization group flow prescription, in the spirit defined in field
theoretical settings, does not exist in this regime.

Let us simplify the formulas further to make a few observations.
Often the metric for spacetimes of interest depends on a single 
``radial'' coordinate, and we choose
the surface $\MM$ to be at constant value with respect to this coordinate. 
Spacetime is therefore parameterized by time, a radial coordinate, and $D-2$
spatial coordinates living on $\MM$.
The integral in~\pref{propt} can then be evaluated, canceling the 
potential divergence
in the numerator, so we write the finite expressions
\bb
c(\tau)=\frac{c_0}{G_D\lk.\sqrt{h} |\theta|^{D-2}\re|_\tau}\ ,
\ee
\bb\label{derprops}
\frac{d}{d\tau}\ln\ c= - (D-2) \frac{R_{ab} n^a n^b}{|\theta |}\ ,
\ee
where $c_0$ is a finite numerical coefficient.
Implied is the statement that for larger values of proper time, 
we penetrate deeper in the bulk and therefore flow to lower energies. This
statement will be made more precise in the next section.
When $|\theta|\rightarrow \infty$, putting an end to the sampling
of the bulk points, and assuming $R_{ab} n^a n^b$ is
finite, equation~\pref{derprops}
indicates that we have reached an infrared fixed point. 
Depending on whether the combination $\sqrt{h} |\theta|^{D-2}$
is finite or infinite, we have a finite or zero central charge.
This implies that all renormalization group flows to lower energies lead to
IR fixed points. This is certainly a desirable statement; we have correlated
caustics with IR fixed points\footnote{
If $R_{ab} n^a n^b$ is to diverge, we expect stringy physics to set
in to regulate the conclusion.
}. 
The other criterion
for the theorem, the null convergence condition, was also a condition
advocated by the work of~\cite{FGPW} 
for the statement of monotonicity. 
Our covariant approach indicates that this observation is a general one.

\section{A covariant UV-IR relation}

We need to prescribe a relation between the proper time and
the energy cutoff in the renormalization group flow. 
We have very little to guide ourselves with in this regard. The UV-IR
relation, as sketched in, for example,~\cite{PEETPOLCH}, 
is a rough scaling relationship; and it
is still associated with various paradoxes. A fundamental formulation
of the relation between scale on the boundary and bulk physics is yet
unknown.
In the spirit of our previous discussion, we should try to write a covariant
UV-IR relation. One can write trivially the statement of~\cite{PEETPOLCH},
\ie\ $g_{tt} dt^2 \sim g_{rr} dr^2$, covariantly
\bb\label{uvir1}
\frac{1}{\mu(\tau)}\simeq \int_{\tau_{UV}}^\tau d\tau'\ n^a \nabla_a t\ .
\ee
Here $t$ is the function foliating the equal time surfaces, $\tau_{UV}$
is the proper time chosen in the UV, and $\mu(\tau)$ is the energy cutoff
associated with the proper time $\tau$. This approach
is particularly naive; and moreover it is not unique. One can
readily write other covariant expressions, particularly
ones that make reference to the second fundamental form and
appear to be more natural choices. Given the correlation
between IR fixed points and caustics we advocated earlier,
it is desirable to state a UV-IR relation such that caustics naturally
correspond to the zero energy limit of the cutoff. 
For the standard scenarios analyzed in the literature
in the context of~\pref{uvir1}, this is certainly the case.
However, our understanding
of the underlying physics in this regard in a general context is
limited; we will then adopt for now equation~\pref{uvir1} as a rough guide 
for the purpose of tracking the scaling relation 
between energy scale and a coordinate in the bulk. 
A more fundamental geometrical
understanding of this issue and the coarse graining prescription
is needed for a more rigorous 
map between the renormalization group and geometry.

\section{Tests and examples}

{\bf Gauged supergravity:}

\vspace{15pt}
We first consider the class of geometries describing domain walls in
gauged supergravities. These solutions interpolate between
two asymptotic AdS regions. For $D=5$, they are believed to correspond to
compactifications
of IIB vacua on manifolds that get deformed from the spherical
geometry of the Freund-Rubin scenario 
as one flows from the UV to the IR. For $D=4$ or $D=7$, they correspond
to M theory compactifications.
A wide class of solutions
were summarized in~\cite{SKENTOWN}, and the metrics have the generic form
\bb
ds^2=e^{2 A(r)} \lk({d{\vec{x}}^2_{(D-2)}}-dt^2\re)+dr^2\ .
\ee
We pick the $D-2$ dimensional manifold $\MM$ as a surface of constant
$r$, and we measure time/energy by the coordinate $t$.
The congruence of in-going null geodesics can easily be found using the 
timelike Killing vector field $\del_t$; one finds
\bb
n^a=\gamma e^{-2 A} \del_t - \gamma e^{-A} \del_r\ ,
\ee
where $\gamma$ is an arbitrary parameter, chosen to be positive, 
that scales the
proper time; this is an arbitrariness characteristic of null geodesics. 
The Christoffel variables are found to be
\bb
\Gamma^r_{ii}= - A' e^{2 A} \eta_{ii}\ \ \ ,\ 
\Gamma^i_{ri}=A'\ ,
\ee
where $\eta_{tt}=-1$ and $\eta_{ii}=+1$ for $i\in \MM$.
We construct the second fundamental form
\bb
B_{bc}=\gamma \lk(
\begin{array}{ccc}
A' e^{-A} & A'     & \cdots 0 \cdots \\
A'        & A' e^A & \cdots 0 \cdots \\
\vdots    & \vdots &                 \\
0         &  0     & -A' e^A \mathbf{1}   \\
\vdots    & \vdots & 
\end{array}
\re)\ .
\ee
The data for the c-function becomes
\bb
\hat{B}=-\gamma A' e^{-A} \mathbf{1}_{D-2}\ \ ,\ \sqrt{h}=e^{(D-2)A}\ ,
\ee
yielding
\bb
c=\frac{c_0}{G_D {A'}^{D-2}}\ .
\ee
This is the expression 
that was proposed for the c-function in~\cite{FGPW,SKENTOWN} along
with compelling evidence in its favor. The form of~\pref{prop}
assures that the parameter $c_0$ here is such that,
in the asymptotic AdS UV region, the central charge is given by that of
the UV fixed point.

\vspace{15pt}
{\bf Dp branes:}

\vspace{15pt}
We next consider the near horizon geometries of $Dp$ branes.
It will be more convenient to first study a class of metrics of the
form
\bb\label{dform}
ds^2=e^{2 A(z)} \lk( f(z)^{-1} dz^2+{d{\vec{x}}^2_{D-2}}-f(z) dt^2 \re)\ .
\ee
We will coordinate transform the $Dp$ brane geometries into this form later.
We choose a constant $z$ and $t$ surface for our manifold $\MM$. The
congruence of null geodesics is given by the vector field
\bb
n^a=\gamma f^{-1} e^{-2A} \del_t + \gamma e^{-2A} \del_z\ .
\ee
The Christoffel variables are
\bbb
\Gamma^z_{zz}&=&\frac{1}{2} \lk( -\frac{f'}{f} + 2 A'\re)\ \ ,\ 
\Gamma^z_{ii}= - f A'\ \ ,\ \nonumber \\
\Gamma^z_{tt}&=& \frac{f}{2} \lk( f' + 2 f A'\re)\ \ ,\
\Gamma^i_{zi}=A'\ \ ,
\Gamma^t_{zt}=\frac{1}{2} \lk( \frac{f'}{f} + 2 A'\re) 
\eee
The data for the c-function becomes
\bb
\hat{B}=-\gamma A' e^{-2A}\ \ ,\ \sqrt{h}=e^{(D-2)A}\ .
\ee
We then get the expression
\bb
c=\frac{c_0}{G_D e^{-(D-2) A} {A'}^{D-2}}\ .
\ee
The function $f(z)$ disappeared from the
expression for the c-function. The role of this function in the metric
is to excite the geometry above extremality, \ie\ to create a thermodynamic
horizon. Correspondingly in the dual description, we excite a finite
temperature vacuum in the same theory. The c-function should not change
when the vacuum reflects a thermodynamic state in the same theory with
the same degrees of freedom. This insensitivity of our expression to
thermodynamic horizons is the second non-trivial piece of evidence
in its favor.
We will come back to this issue later in the context of the AdS
Schwarzschild black hole; for now,
let us proceed to $Dp$ branes.

A $Dp$ brane metric in the 
Maldacena scaling regime is given by~\cite{HORSTROM,MALDA2}
\bb
ds_{Str}^2=\lk(\frac{r}{q}\re)^{\frac{7-p}{2}}
\lk( {d{\vec{x}}^2_{(p)}} - f dt^2\re)
+\lk(\frac{q}{r}\re)^{\frac{7-p}{2}}
\lk( f^{-1} dr^2 + r^2 d\Omega_{8-p}^2\re)\ .
\ee
with the dilaton being
\bb
e^\phi= \lk(\frac{q}{r}\re)^{\frac{(7-p)(3-p)}{4}}\ ,
\ee
and
\bb
q^{7-p} \simeq \gs N\ \ ,\ f=1-\lk(\frac{r_0}{r}\re)^{7-p}\ ,
\ee
where ({\em after} taking the decoupling limit!) we have chosen units such that
$\alp=1$. Energy in the SYM dual is measured with respect to the coordinate
time $t$.
Given that we lack numerical accuracy
in the relation between energy scale 
and radial extent $r$ at present (for that matter, we also lack 
rigorous conceptual 
understanding in this context), we will now start being careless with numerical 
coefficients and aim at determining only the scaling of the c-function
with respect to the physical parameters of the SYM theory.
We apply the coordinate transformation
\bb
r=z^\frac{2}{p-5}\ \ \ \mbox{  with  }\ p<5\ .
\ee
After rescaling the metric to the Einstein frame, as well as absorbing
certain constants in the transverse coordinates
\bb\label{rescale}
g_{\mu\nu}^{Ein}=e^{-\phi/2} g_{\mu\nu}^{Str}\ \ ,\ 
(p-5)\frac{q^{\frac{p-7}{2}}}{2}\lk(\vec{x},t\re)\rightarrow
\lk(\vec{x},t\re)\ ,
\ee
we get
\bb\label{dbmetric}
ds_{Ein}^2=q^{\frac{(p+1)(7-p)}{8}} z^{\frac{(p-3)^2}{4(p-5)}}
\lk(\frac{2}{p-5}\re)^2
\lk(\frac{1}{z^2} \lk( f^{-1} dz^2 - f dt^2 + {d{\vec{x}}^2_p}\re)
+ \lk(\frac{p-5}{2}\re)^2 d\Omega_{8-p}^2\re)\ .
\ee
Note that we have effectively rescaled the SYM energy. 
This metric is of the form~\pref{dform}
except for the transverse $8-p$ sphere factor. Even though
it is straightforward to extend our formalism to this
extended space with the transverse sphere, it is easier to track
the scaling of the physical parameters by imagining that we have compactified
the geometry on this sphere, with the effect that the gravitational
coupling in the lower $D=p+2$ dimensions scales as
\bb
G_{(p+2)}\simeq \frac{\gs^2}{Vol(\Omega_{8-p})}=
\gs^2 z^{p-8} e^{(p-8) A(z)}\ ,
\ee
where $A(z)$ refers to the corresponding function identified from 
matching~\pref{dform} with~\pref{dbmetric}.
We then have
\bb
A'\simeq \frac{1}{z}\ .
\ee
The c-function becomes
\bb
c=c_0 \frac{z^{8-p} e^{8 A}}{\gs^2 {A'}^p}
=c'_0 \gs^{\frac{p-3}{2}} N^{\frac{p+1}{2}} z^{\frac{(p-3)^2}{p-5}}\ ,
\ee
where $c_0$ and $c'_0$ are numerical coefficients.
Applying the UV-IR relation given by~\pref{uvir1}
(necessarily at zero temperature; see comments 
clarifying the relevance of this statement in the
Schwarzschild black hole section), we get~\cite{PEETPOLCH}
\bb
\mu(z)\sim \frac{1}{\sqrt{\gs N} z}\ ,
\ee
where we interpret $\mu$ as the renormalization energy cutoff scale.
We have eliminated the proper time in favor of the coordinate $z$ using
the trajectory of the geodesics.
We have also undone the rescaling of the time variable
in~\pref{rescale}, so that $\mu$ is
energy scale as measured in the SYM theory.
Putting things together, and defining the effective large $N$
dimensionless coupling as
\bb
\geff^2(\mu)\equiv \gym^2 N \mu^{p-3}\ ,
\ee
where $\gym^2=\gs$,
we arrive at an expression for the c-function for the $p+1$ dimensional
SYM theory
\bb\label{cDpmain}
c_{Dp}(\mu)\simeq \gs^{\frac{p-3}{5-p}} N^{\frac{p-7}{p-5}}
\mu^{\frac{(p-3)^2}{5-p}}\simeq
\geff^{2 \frac{p-3}{5-p}}(\mu) N^2\simeq c_{SYM}(\geff)\ .
\ee
The first thing to
note is that, when the energy scale is $\mu\sim \mu_{YM}$ such that the
curvature scale in the region of space where the integral of
equation~\pref{propt} is evaluated becomes of order the string scale,
we have $\geff^2(\mu_{YM}) \sim 1$, and therefore, for all $p$,
\bb
c_{Dp}(\mu_{YM})\simeq N^2\ ,
\ee
\ie\ the gravitational description breaks down at the
Horowitz-Polchinski correspondence point~\cite{CORR1} and
the central charge scales as in the perturbative SYM regime.
This happens in the UV for $p<3$ and in the IR for $p=4$. For $p=3$,
we note that the c-function is constant and of order $N^2$ as expected
for the conformal $3+1$d $\NN=4$ SYM theory. 

Now let us analyze the different
scenarios more closely: for $p=1$, we get
\bb\label{cd1}
c_{D1}(\mu) \simeq \frac{N^2}{\geff(\mu)} \simeq \frac{N^{3/2}}{\gym} \mu\ .
\ee
This result was obtained in~\cite{HASHITZHA} by different methods; 
the authors there
could use the correlation function with insertions of
two energy-momentum tensors to read off the central charge.
This c-function, as noted by them, interpolates between
the 1+1 dimensional SYM and the Matrix String regimes; the latter
arises in the IR at energy scales $\mu_{MS}\sim \gym/N^{1/2}$,
which is again a Horowitz-Polchinski correspondence point in a dual
geometry~\cite{MSSYM123}; equation~\pref{cd1} yields at this energy scale 
\bb
c_{D1}(\mu_{MS})\sim N
\ee
as expected for the Matrix String~\cite{MOTL,DVV}.

Next, consider $p=2$; we have
\bb
c_{D2}(\mu) \simeq \frac{N^2}{\geff^{2/3}(\mu)} \simeq 
\frac{N^{5/3}}{\gym^{2/3}}\mu^{1/3}\ .
\ee
Moving from the $2+1$ SYM theory to the IR, at energy scales
$\mu_{M2}\sim \gym^2 N^{-1/2}$, 
as shown in~\cite{MSSYM123}, the membrane theory is encountered.
We find the corresponding central charge is
\bb
c_{D2}(\mu_{M2})\sim N^{3/2}\ .
\ee
This is indeed the proposed behavior for the membrane theory~\cite{GUBKLEB}.
Beyond this point, the geometry becomes $AdS_4$, and it appears
we have reached a fixed point of flow\footnote{
One may expect naively that the energy scale $\mu_{M2}$ 
must be $1/R_{11}$; however,
the geometry of the lifted $D2$ branes is given by that of smeared $M2$ 
branes, whose near horizon geometry is {\em not} $AdS_4$; the energy
scale $\mu_{M2}\sim \gym^2 N^{-1/2}$ was identified in~\cite{MSSYM123}
as the scale where the localized membrane theory sets in; it is this
geometry which is $AdS_4$. 
Correspondingly, we find the characteristic $N^{3/2}$ scaling
for membranes at this scale of the flow.
}.

For $p=4$, we flow from the SYM in the IR to the $(2,0)$ theory
on a circle sitting in the UV. We have
\bb\label{n3}
c_{D4}(\mu)\sim \gym^2 \mu N^3\ .
\ee
As we flow to the UV,
we will start probing the size of the eleventh dimension. 
This happens at $\mu_{M5}\sim 1/R_{11}\sim \gym^{-2}$.
The central charge then becomes
\bb
c_{D4}(\mu_{M5})\sim N^3.
\ee
This is indeed the characteristic scaling for the
central charge of the M5 branes~\cite{HENNSKEN}. The geometry becomes
beyond this scale $AdS_7\times S^4$, 
\ie\ the near horizon geometry of M5 branes.

The reader may have noticed that $p=5$ was a special case in our
analysis. The coordinate transformation applied for these
examples breaks down in this setting. In this case,
one probes the delicate NS5 brane 
geometry; a more careful analysis of the geodesic flow is in order.
The results are bound to have more of a
predictive nature than of a test of our proposal; we will postpone this 
task to the future.

Equation~\pref{cDpmain} appears then to correctly reflect the renormalization
group flow in $p+1$ dimensional SYM theories for $p<5$. Furthermore,
the expression is insensitive to excitations of the vacuum above extremality
to finite temperatures. We will say more on this issue in the next section.
One may argue that the matching onto the M2 and
M5 brane central charges is not terribly impressive since the geometries
become AdS at these energy scales, and our expression is tuned
to give the right answer for AdS spaces. 
We note however that, 
at the Horowitz-Polchinski correspondence points, this issue cannot be raised.
We have two powers of the dimensionless quantities
$\geff$ and $N$ to check against; 
if one gets fixed by the AdS region, the other is free. Note that in
the case for $D1$ branes, the Horowitz-Polchinski correspondence criteria
bound both sides of the flow. And these matched the proper asymptotics
known from other reliable methods. Turning around the argument,
this becomes a test
of our hypothesis that the c-function is expressible in terms
of local geometrical quantities; a statement which is intuitively
in tune with the prescription of the renormalization group flow.
When we reach a fixed point, \ie\ an AdS region, it is irrelevant
how we got there; there may be different routes to flowing to a fixed
points from different neighboring conformal field theories.
The outcome must be the same; the central charge is fixed by the end point,
\ie\ by the cosmological scale and gravitational coupling of the
AdS region. This viewpoint, along with the non-trivial matchings
with the perturbative SYM regimes for all $p$ and the Matrix String
theory, constitutes compelling evidence in favor of a local expression
for the central charge.

\vspace{15pt}
{\bf AdS Schwarzschild:}

\vspace{15pt}
We briefly explore here the four dimensional AdS Schwarzschild black
hole case to illustrate a previous point in a simpler setting. 
Let us consider the metric~\cite{WITPHASE}
\bb\label{bh}
ds^2=-g(r) dt^2 + f(r) dr^2 + r^2 d\theta^2 + r^2 \sin^2 \theta\ d\phi^2\ .
\ee
For the AdS black hole, we have
\bb
g(r)=1-\frac{r_0}{r} + \frac{r^2}{l^2}=f(r)^{-1}\ ,
\ee
where $l$ is related to the cosmological constant. One then finds
\bb
c\sim f(r)\ g(r)\ .
\ee
The inverse relation between $f$ and $g$ characteristic
of horizon excitations yields a constant c-function. As in
the near extremal $Dp$ brane cases, we find here as well that our assessment of
the central charge in the theory does not get affected by a thermodynamic
horizon. The central charge for the AdS Schwarzschild black hole
geometry is a constant equal to the value set by the asymptotic AdS region.

Note that this example also demonstrates that
the disappearance of the function $f$ from the central charge in the
cases of near-extremal $Dp$ branes was {\em not} a result of focusing on the
space $\MM$ transverse to the $z-t$ plane; it is the relative relationship
between $f$ and $g$ that the central charge probes. One may get troubled 
from the fact that
we used the {\em zero temperature} UV-IR relation when we wrote the
central charge as a function of the cutoff energy; \ie\ we used
the extremal metric. The complaint would be that the insensitivity of
the central charge, written as a function of energy scale, to 
the presence of the
black horizon was partly put by hand. The point is that the relation
between energy scale in the boundary theory and extent in the bulk 
should be an {\em independent} statement; a covariant formulation
of the UV-IR statement should be insensitive to the presence of a
thermodynamic horizon in the geometry independently from any other statement.
It would be unphysical if the presence of a 
background thermal bath affected
our assessment of the relation between the location of an excitation
in the bulk and its energy as measured in the boundary theory.
This would have been needed if the 
cancelation of the horizon factors in the central
charge expression was not to occur. This feature of our expression
is then positive evidence in its favor.

In what sense then is the presence of a black horizon special?
It seems that we are drawn to the conclusion that the central charge 
for the AdS Schwarzschild black hole geometry is set
by the asymptotic AdS region; the black hole is simply a thermal
state in the conformal field theory dual to the AdS vacuum~\cite{BDHM,BKLT}. 
The answer has to do with the causal aspect of the horizon. 
Consider a surface $\MM$ sitting {\em at} the horizon. As
Bousso notes, there are now three classes of null congruences which
are candidates for sampling the bulk. One set will sample the inside
of the horizon, but the other two sets, the 
trapped geodesics, will sample
the surface area. These ones saturate the sampling criterion
$\theta \le 0$; \ie\ they satisfy $\theta=0$. 
This indicates that the {\em gravitational} dynamics 
{\em at} the horizon in some sense is a holographic dual to 
the gravitational dynamics {\em within} the horizon; 
both descriptions being duals to a conformal field theory.
This viewpoint, in its current fetal state,
presents an intriguing
marriage between our understanding of the special causal aspects
of a black horizon, its thermodynamic character, and renormalization
group flow.

In principle, one can arrange matter configurations
so as to curve spacetime 
as in~\pref{bh} with arbitrary $f$ and $g$. Asymptotically, we must recover
AdS space with the cosmological constant set by our gravitational
action. Our analysis suggests that, if this setup is stable, unlike
the black hole scenario, it would correspond
to perturbing the boundary theory away from conformality,
generating non-trivial renormalization group flow.

The generalization of this example to higher dimensions is
straightforward; we expect no change in the conclusions.

\section{Discussion}

Let us recap the proposal and critically assess the evidence we have
presented in its favor.
We used a principle of covariant holography
and a set of intuitively driven,
yet non-rigorous, arguments to define holographic duals. In this context,
we proposed a c-function for the boundary theory; it is a
geometrical, local, covariant expression holographic in nature
\bb\label{finalform}
c\sim \frac{1}{\int_\MM \det\ \hat{B}}\ .
\ee
The inverse central charge is simply written as the integral of the 
determinant of the second characteristic form on $\MM$.
The expression was explicitly tuned to yield a constant for AdS spaces.
Unfortunately , we do not 
have a more physical understanding of the form of~\pref{finalform}. 
If this proposal is indeed correct, it is a statement about understanding
the holographic encoding of information in the language of
the renormalization group.

The evidences we presented in favor of~\pref{finalform} were as follows:

\begin{itemize}
\item We could prove a c-theorem: this essentially followed from
Raychaudhuri's equation.
One of the two criteria for a monotonically decreasing
c-function, the condition $\theta<0$, 
correlates with Bousso's criterion for sampling the bulk
space for information; the second criterion, 
the null convergence condition, was advocated independently in the
example of~\cite{FGPW}.

\item For domain wall solutions, our result agrees with~\cite{FGPW,SKENTOWN}.
This may be regarded as merely a test for the possibility to formulate
a c-function through the
formalism of congruences of null geodesics.

\item For the $Dp$ brane geometries, our expression appears to
interpolate correctly between known asymptotics. This constitutes 
a non-trivial test for the principle that a covariant and local expression
for the c-function exists.
As such however, any other covariant local expression
is a candidate as well.

\item Exciting a thermodynamic horizon in the bulk space does not
change the central charge. This constitutes a non-trivial test for the form
we have proposed, beyond the test for covariance. 

\end{itemize}

The proposal~\pref{finalform} may be
incomplete, with additional corrections
needed as we move further into the bulk~\cite{BALASTRESS}. 
The successes we demonstrated
may have been accidents due to certain symmetries in the cases
considered. Any covariant term, invariant under boundary diffeomorphisms,
and vanishing in the maximally symmetric AdS case is a priori allowed.
Less interestingly, terms involving 
arbitrary powers of $\det \hat{B}$ and $\theta$
such that the dimensions are right are allowed;
we can write combinations involving objects like 
$\mbox{Tr } \hat{B} \ldots \hat{B}$.
Such possibilities are endless, as well as being uninteresting;
unfortunately, without examples that probe the effect of shear,
one cannot distinguish between them.
We propose the expression~\pref{finalform} for the c-function
as it appears to be the
simplest and most natural form amongst these possibilities.  
On the other hand, more general
terms can be multiplied by powers of the gravitational coupling
to make them dimensionless; their origin would then 
probably be stringy. One may in principle add terms constructed
from pull-backs of the curvature tensors; adding Weyl tensor
dependent terms would not affect our conclusions in the cases of
extremal $Dp$ branes and domain walls; it would however
change the conclusion for near extremal $Dp$ branes, which would be
undesirable.  It is possible that our approach may be a first
order approximation to the underlying physics; perhaps probing the geometry
by geodesics can go so far; one may need to study the full quantum field theory
in a given background geometry (or for that matter the full string theory)
to decode renormalization group data from gravitational physics. 
On the other hand, the principle of covariant holography accords
an attractive special physical role to null geodesics\footnote{
Null geodesics were also used in a similar approach
in the work of~\cite{AG}.}.
We certainly expect corrections of string theoretical origin
as the geometrical description starts to break down. 
However, within energy scales where the low energy gravity sector is a good
approximation, a fundamental quantity like the c-function may be expected
to have a simple geometrical representation such as~\pref{finalform}. 
This is in the spirit of the
frugal statement that relates entropy of a black hole with the area of its
horizon. We believe that
we have presented enough evidence to make the proposal
worthy of further investigation.

One of the most attractive aspects of~\pref{finalform} is the fact
that it is in practice easy to computationally handle.
It can readily be applied to a myriad of geometries, tested, as well as
used to understand the nature of certain ill-understood dual
theories (such as five branes). 
There are also more stringent tests that 
the expression can be subjected to:
in particular, an understanding of the relation between the
first order Callan-Symanzik equations and the second order Einstein
equations is of direct relevance to this proposal. 
Work in this direction is in progress.

We stated in the beginning of the first section the condition that
the gravitational vacuum under consideration should solve Einstein's
equations in the presence of a negative cosmological constant.
From the string theory side, we know that there exists an energy regime
that screens out regions of spacetime that are not candidates
for holography. This typically leads to focusing on the near horizon 
geometries of $Dp$ branes, which are conformal to 
AdS spaces. On the side of the boundary theory, fixed points
play a fundamental role in defining renormalization group flows. These
special points indeed correspond to AdS spaces. It is in this light
that we are motivated to state that holography in general, and the
formalism we presented in particular, need to be thought of in the context
of a gravitational theory with a negative cosmological constant.
This line of thought rules out extending these ideas to flat Minkowski
space. It would be interesting however to explore this approach
in scenarios where the spacetime does not admit a timelike Killing
vector field.

An important issue that we have not been able to
address properly is a covariant formulation of the UV-IR
correspondence. This issue is related to an understanding of the
process of coarse graining as seen by the gravity side. 
A possible picture for this was presented in~\cite{BALARG}. On
the other hand, it is
tempting to believe that the gravitational vacuum that solves
Einstein's equations reflects the state of the dual
theory at all energy scales; locally, foliations are snapshots
of the theory at different energy scales. The metric and its first
derivative on $\MM$ (essentially the content of the congruence data) encode
all the necessary information about the theory at a given scale. 
From the difficulty we are having to formulate a covariant UV-IR
correspondence, it appears that this line of thought may be only
part of the whole picture.

Finally, certain simplifying assumptions were made in the text to
arrive at leaner conclusions and to focus on the relevant physics.
The assumption of shearless flow however may hold rich physics. 
Even though it may appear straightforward to generalize the approach to
this case, there are subtleties which we do not understand in this
context. Flow with shear is in particular a characteristic of
boundary theories coupled to background gravity. Our understanding
of the effect of this on the renormalization prescription is limited.

If a fundamental relation between renormalization group flow and geometry
exists, it should be possible 
to find a geometrical interpretation for every object in the
renormalization group prescription. We hope to have stimulated 
further investigation in this direction.

\vspace{24pt}
\paragraph{\bf Acknowledgments:}
I have enjoyed conversations with V. Balasubramanian, E. Flanagan,
H. Tye, K. Narayan, and S. Teukolsky. I am grateful to K. Narayan
for a careful reading of a draft.
The comments about possible corrective terms to
our proposal arose in discussions with
V. Balasubramanian and H. Tye.
This work was supported by NSF grant PHY-9513717.

\providecommand{\href}[2]{#2}\begingroup\raggedright\endgroup

\end{document}